%% file: ms.tex
\documentclass{emulateapj}
\usepackage{apjfonts}

\newcommand{\Dnu}{\mbox{$\Delta \nu$}}
\newcommand{\acena}{\mbox{$\alpha$~Cen~A}}
\newcommand{\acenb}{\mbox{$\alpha$~Cen~B}}
\newcommand{\acen}{\mbox{$\alpha$~Cen}}
\newcommand{\bhyi}{\mbox{$\beta$~Hyi}}

\newcommand{\eboo}{\mbox{$\eta$~Boo}}

\newcommand{\muHz}{\mbox{$\mu$Hz}}

\newcommand{\new}[1]{{\bf #1}}
\renewcommand{\new}[1]{{#1}}
\newcommand{\myomit}[1]{{\bf Omit: #1}}
\renewcommand{\myomit}[1]{\relax}

\slugcomment{To appear in ApJ Letters}

\shorttitle{Near-surface corrections for stellar oscillations}
\shortauthors{Kjeldsen et al.}

\begin{document}

\title{Correcting stellar oscillation frequencies for near-surface effects}
 
\author{
Hans~Kjeldsen\altaffilmark{1},
Timothy R. Bedding\altaffilmark{2} and
J{\o}rgen Christensen-Dalsgaard\altaffilmark{1}
}

\altaffiltext{1}{Danish AsteroSeismology Centre (DASC), Department of
Physics and Astronomy, University of Aarhus, DK-8000 Aarhus C, Denmark;
hans@phys.au.dk, jcd@phys.au.dk}

\altaffiltext{2}{Institute of Astronomy, School of Physics, University of
Sydney, NSW 2006, Australia; bedding@physics.usyd.edu.au}

\begin{abstract} 
In helioseismology, there is a well-known offset between observed and
computed oscillation frequencies.  This offset is known to arise from
improper modeling of the near-surface layers of the Sun, and a similar
effect must occur for models of other stars.  Such an effect impedes
progress in asteroseismology, which involves comparing observed oscillation
frequencies with those calculated from theoretical models.  Here, we use
data for the Sun to derive an empirical correction for the near-surface
offset, which we then apply three other stars (\acena{}, \acenb{} and
\bhyi).  The method appears to give good results, in particular providing
an accurate estimate of the mean density of each star.
\end{abstract}


\section{Introduction}

Both helio- and asteroseismology involve comparing observed oscillation
frequencies with those calculated from theoretical models.  However, for
the Sun there is a long-standing systematic offset between observed and
computed frequencies that is known to arise from improper modeling of the
near-surface layers \citep{ChDDL88,DPV88,ChDDA96,ChD+T97}.  This offset is
independent of the angular degree of the mode ($l$) and increases with
frequency.  A similar offset must occur for models of other stars, and
should be taken into account whenever observations of stellar oscillations
are compared with theory.  In this Letter we use data for the Sun to derive
an empirical correction for these near-surface effects and show how to
apply this to other stars.  The method appears to give good results, in
particular providing an accurate estimate of the mean density of each star.

\section{Method}

The p-mode oscillations in solar-type stars for a given angular degree~$l$
are approximately equally spaced in frequency, with a separation of $\Dnu$
(the so-called large separation; see \citealt{ChD2004} for a review of the
theory of solar-like oscillations).  Since the offset from incorrect
modelling of the near-surface layers is independent of~$l$, we can derive
the correction by considering only the radial modes ($l=0$) and then apply
it to all modes. 

Suppose we have a set of observed frequencies for radial modes, $\nu_{\rm
obs}(n)$, where $n$ is the radial order.  Suppose also that $\nu_{\rm best}(n)$
are the frequencies from the best model, by which we mean the one
that best describes the parameters and internal structure of the star, but
which still fails to model correctly the surface layers,

For the Sun, the difference between observed and best model frequencies
turns out to be well fitted by a power law, \new{which has the convenient
property of being free of a frequency scale (see also \citealt{ChD+G80})}:
\begin{equation}
  \nu_{\rm obs}(n) - \nu_{\rm best}(n) = a \left(\frac{\nu_{\rm
  obs}(n)}{\nu_0}\right)^b, \label{eq.fobs-fref}
\end{equation}
where $\nu_0$ is a suitably chosen reference frequency, and $a$ and $b$ are
parameters to be determined.  

Suppose further that we have calculated a reference model that has
frequencies $\nu_{\rm ref}(n)$ and is close to the best model.  From homology
scaling it then follows that, to a good approximation,
\begin{equation}
  \nu_{\rm best}(n) = r \nu_{\rm ref}(n),\label{eq.fbest.fref}
\end{equation}
where the scaling factor $r$ is related to the mean densities
$\bar \rho_{\rm best}$ and $\bar \rho_{\rm ref}$ of the best and reference
models by
\begin{equation}
  \bar{\rho}_{\rm best} = r^2 \bar{\rho}_{\rm ref}.  \label{eq.rho}
\end{equation}
Given a determination of $r$, equation~(\ref{eq.rho}) provides our best
estimate of the mean density of the star.

Substituting equation~(\ref{eq.fbest.fref}) into
equation~(\ref{eq.fobs-fref}) gives
\begin{equation}
  \nu_{\rm obs}(n) - r \nu_{\rm ref}(n) = a \left(\frac{\nu_{\rm
  obs}(n)}{\nu_0}\right)^b \label{eq.fobs-rfref} 
\end{equation}
and differentiating with respect to $n$ gives
\begin{equation}
  \Delta\nu_{\rm obs}(n) - r \Delta\nu_{\rm ref}(n) = a b \left(\frac{\nu_{\rm
            obs}(n)}{\nu_0}\right)^{b-1} \frac{\Delta\nu_{\rm
      obs}(n)}{\nu_0}.\label{eq.Deltanu}
\end{equation}
Combining and rearranging these last two equations gives
\begin{equation}
  r = (b-1)\left(b \frac{\nu_{\rm ref}(n)}{\nu_{\rm
	obs}(n)} - \frac{\Delta\nu_{\rm ref}(n)}{\Delta\nu_{\rm
	obs}(n)}\right)^{-1} \label{eq.r}
\end{equation}
and
\begin{equation}
  b = \left(r \frac{\Delta\nu_{\rm ref}(n)}{\Delta\nu_{\rm
      obs}(n)} - 1\right) 
      \left(r \frac{\nu_{\rm ref}(n)}{\nu_{\rm
	obs}(n)} - 1\right)^{-1}. \label{eq.b}
\end{equation}
If we know $b$ then we can calculate $r$ using equation~(\ref{eq.r}), or
{\em vice versa\/} using equation~(\ref{eq.b}).  We can then obtain~$a$
using equation~(\ref{eq.fobs-rfref}).

We now show how to apply the above method to a set of observed and
calculated frequencies.  Suppose we have frequencies for $N$ radial modes
with orders $n_1, n_2, \ldots, n_N$ (not necessarily consecutive).  We use
these to calculate the four terms needed to evaluate equation~(\ref{eq.r})
or~(\ref{eq.b}).  For $\nu_{\rm obs}(n)$ and $\nu_{\rm ref}(n)$, we simply use
the means of the given sets of frequencies, which we denote by $\langle
\nu_{\rm obs}(n) \rangle$ and $\langle \nu_{\rm ref}(n) \rangle$.

To estimate the large separations, we calculate the slope of a linear
least-squares fit to the given frequencies (as a function of~$n$):
\begin{eqnarray}
  \langle \Delta\nu_{\rm obs}(n) \rangle &=& 
  \frac{\sum_{i=1}^{N} 
    \left(\nu_{\rm obs}(n_i) - \langle \nu_{\rm obs}(n)\rangle\right)
    \left(n_i-\langle n\rangle\right)}
  {\sum_{i=1}^{N} \left(n_i-\langle n\rangle\right)^2}\\
  \langle \Delta\nu_{\rm ref}(n) \rangle &=& 
  \frac{\sum_{i=1}^{N} 
    \left(\nu_{\rm ref}(n_i) - \langle \nu_{\rm ref}(n)\rangle\right)
    \left(n_i-\langle n\rangle\right)}
  {\sum_{i=1}^{N} \left(n_i-\langle n\rangle\right)^2}.
\end{eqnarray}
We must then assume a value for either $b$ or $r$, and use
equation~(\ref{eq.r}) or~(\ref{eq.b}) to estimate the other.  Finally, the
value of~$a$ is found from equation~(\ref{eq.fobs-rfref}), as follows:
\begin{equation}
  a = \frac{\langle \nu_{\rm obs}(n) \rangle - r \langle \nu_{\rm ref}(n) \rangle}
      {N^{-1} \sum_{i=1}^{N} \left(\nu_{\rm obs}(n_i)/\nu_0\right)^b}.
\end{equation}

We now proceed to apply this method to the Sun in order to measure~$b$
(\S\ref{sec.sun}), and then adopt this value of~$b$ for other stars
(\S\ref{sec.stars}).

\section{Application to the Sun}\label{sec.sun}
   
For the Sun we took Model~S of \citet{ChDDA96}, as listed in the first row
of Table~\ref{tab.models}.  We assumed this to be the ``best'' solar model,
in the sense defined above, which means we can set $r=1$ (see
equation~\ref{eq.fbest.fref}).  For the observed solar frequencies, we used
those measured by \citet{LBB97} with the GOLF instrument on the SOHO
spacecraft.

\begin{figure}
\epsscale{0.9}
\plotone{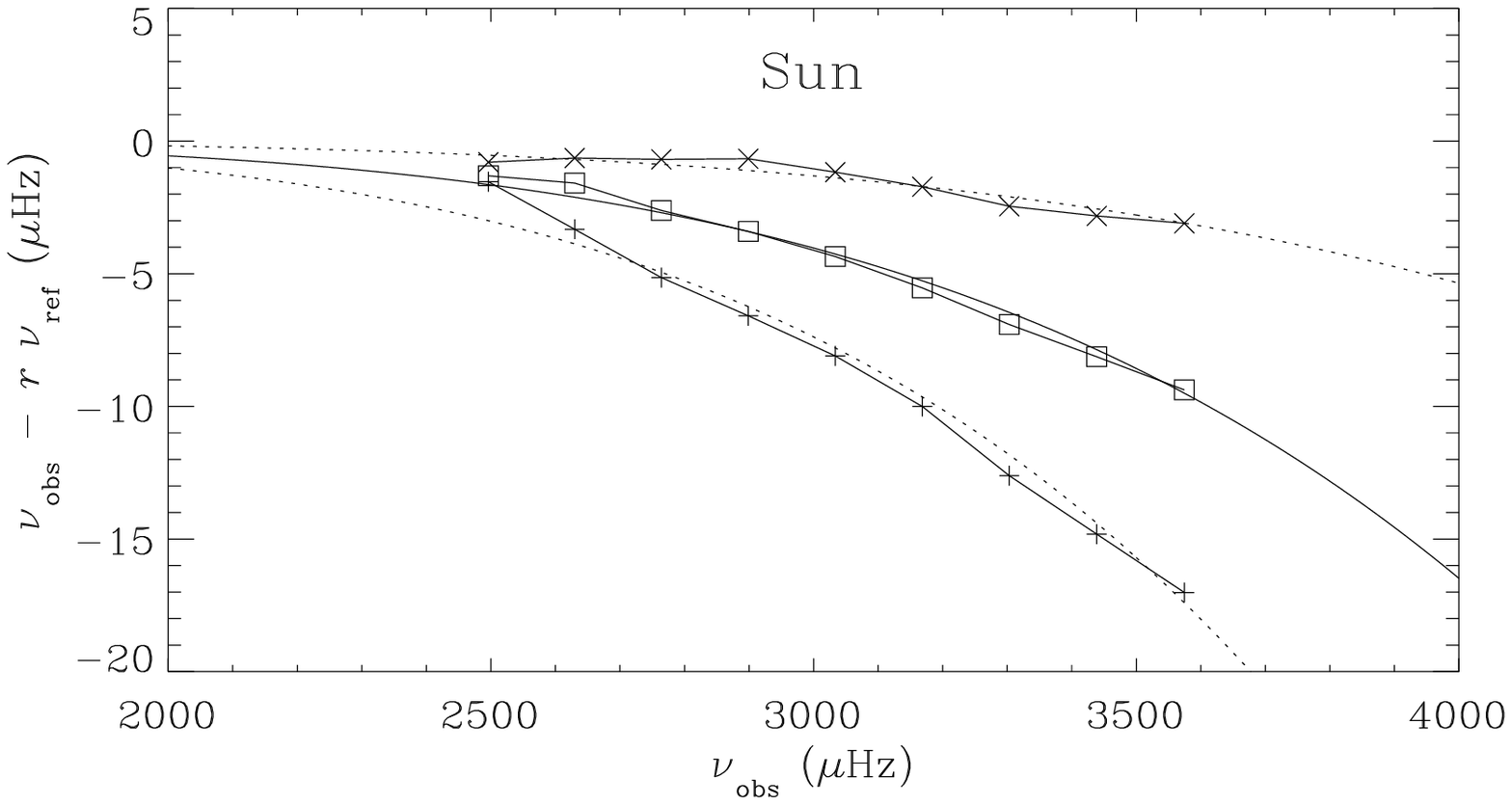}
\caption[]{\label{fig.solar} The difference between observed and calculated
frequencies for radial modes in the Sun.  The squares are for Model~S, with
the solid curve showing a fit to equation~(\ref{eq.fobs-rfref}) with $r=1$,
which gives $b=4.90$ (see~\S\ref{sec.sun}).  Also shown are the results of
applying the same corrections to Model~S$^-$ (crosses) and Model~S$^+$
(pluses).  The dotted curves show the corrections calculated from
equation~(\ref{eq.fobs-rfref}).}
\end{figure}

We followed the procedure described above, choosing $\nu_0 = 3100\,\muHz$ and
setting $r=1$, and using the data to measure~$b$ and~$a$.  We have
chosen to use the nine modes centred at the peak of the oscillation power,
from which we obtained the results shown in the first line of
Table~\ref{tab.calcs}, and a value of $b=4.90$.  The differences between
observed and Model~S frequencies are plotted as the squares in
Fig.~\ref{fig.solar}, and the solid curve is the function given by
equation~(\ref{eq.fobs-fref}).

\new{ The above fit was made for the strongest 9 radial modes in the Sun.
We repeated the analysis for different numbers of modes (all values from 7
to 13), and found the derived value of $b$ to range from 4.4 to 5.25.
Clearly, the frequency differences do not exactly follow a power law, and
so the exponent in the power-law fit depends substantially on the frequency
range.  Importantly, the value for $a$ varies by less than 0.1\,\muHz{} in
all cases.  }

In addition to Model~S, we also considered models denoted~S$^-$ and~S$^+$
from the same evolution sequence as Model~S, but with ages of 2.25 and
7.44\,Gyr, respectively.  The parameters of these models are given in
Table~\ref{tab.models}.  We kept $b$ fixed at the value found for the
``best'' model ($4.90$) and used equation~(\ref{eq.r}) to estimate $r$.
The results are shown in Table~\ref{tab.calcs}, and also in
Fig.~\ref{fig.solar}.  Importantly, the derived density of the Sun ($r^2
\bar{\rho}_{\rm ref}$) is correct for both these calculations, despite the
very different densities of the models themselves, giving us confidence
that the method has been successful.  \new{Even more importantly, the
derived density of the Sun is completely insensitive to the choice of $b$.
This reflects the fact that the value of~$r$ obtained by fitting to
equation~\ref{eq.fobs-rfref} is not sensitive to the exact form of the
function on the right-hand side, so long as that function tends to zero
with decreasing frequency.}

\section{Application to \acena, \acenb{} and \bhyi}\label{sec.stars}

We have considered three stars for which observations and models have been
published.  For \acena{} we took observed frequencies (radial modes only)
from four sources: \citet{B+C2002}, \citet{BKB2004}, \citet{BazBK2007} and
\citet{FCE2006}.  This gave a set of 33 measured frequencies for 11 orders,
all of which were given equal weight in the fitting process.  For \acenb{}
the observed frequencies were taken from \citet{KBB2005} and those for
\bhyi{} from \citet{BKA2007}.  \new{In each case, we used all the
detected $l=0$ frequencies.}

The models that we have used for these stars are listed in
Table~\ref{tab.models}.  These include published models of \acena{} and~B
by \citet{MPL2000} and \citet{TPM2002}, \new{of \acenb{} by
\citet{TBG2008},} and of \bhyi{} by \citet{F+M2003}.  In addition to
published models, we have considered several computed with the Aarhus
stellar evolution code (ASTEC, \citealt{ChD2008}).  Models~A and~B are
models of the \acen{} system, computed with essentially the same physics as
Model~S and fitted (T. Teixeira et al.\ 2008, in preparation) to the
observations of \citet{BKB2004} and \citet{KBB2005}.  For \bhyi, Model~H
matches the parameters reported by \citet{NDB2007} and was computed with
similar physics, but neglecting diffusion and settling.  Finally,
Models~H$^-$ and~H$^+$ are from the same evolution sequence as Model~H, but
with substantially different ages and densities, which bracket those of
Model~H.

\begin{figure}
\epsscale{0.9}
\plotone{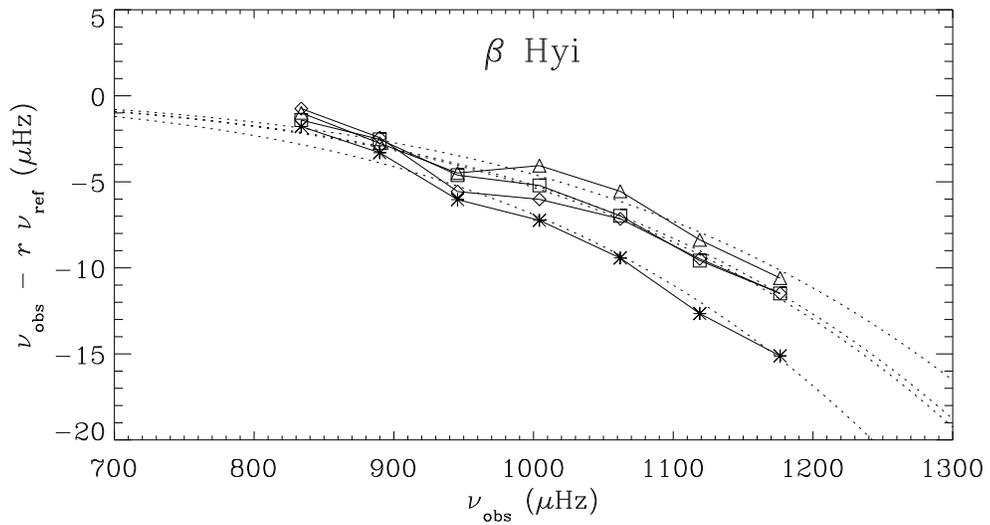}
\caption[]{\label{fig.bhyi} The difference between observed and calculated
frequencies for radial modes in \bhyi.  The models shown are: Model~H
(squares), Model~H$^-$ (triangles), Model~H$^+$ (diamonds) and FM2003
(asterisks). The dotted curves show the corrections calculated from
equation~(\ref{eq.fobs-rfref}).}
\end{figure}

\begin{figure}
\epsscale{0.9}
\plotone{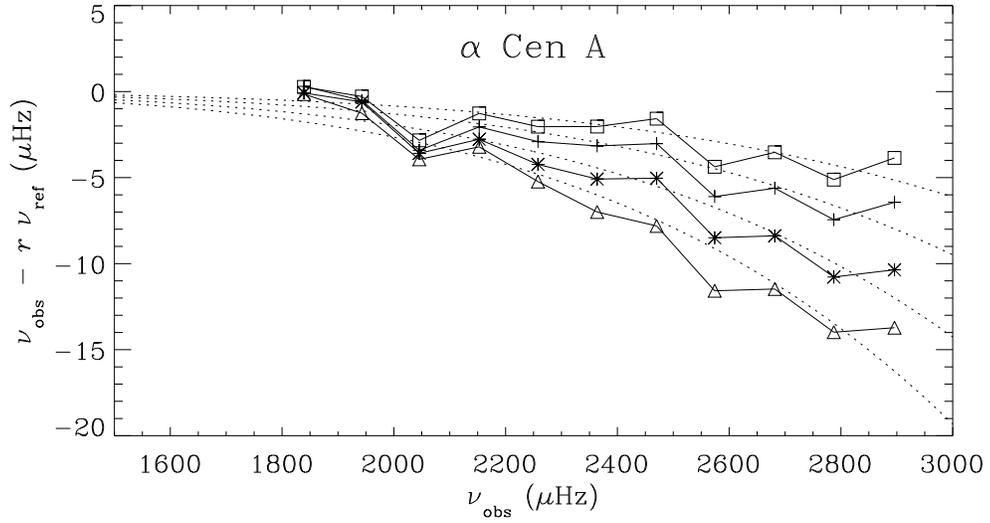}
\caption[]{\label{fig.acena} Same as Fig.~\ref{fig.bhyi}, but for \acena.
The models shown are: Model~A (squares), M2000~A (triangles), Model~S$^+$
(pluses) and Th2002~A (asterisks).}
\end{figure}

\begin{figure}
\epsscale{0.9}
\plotone{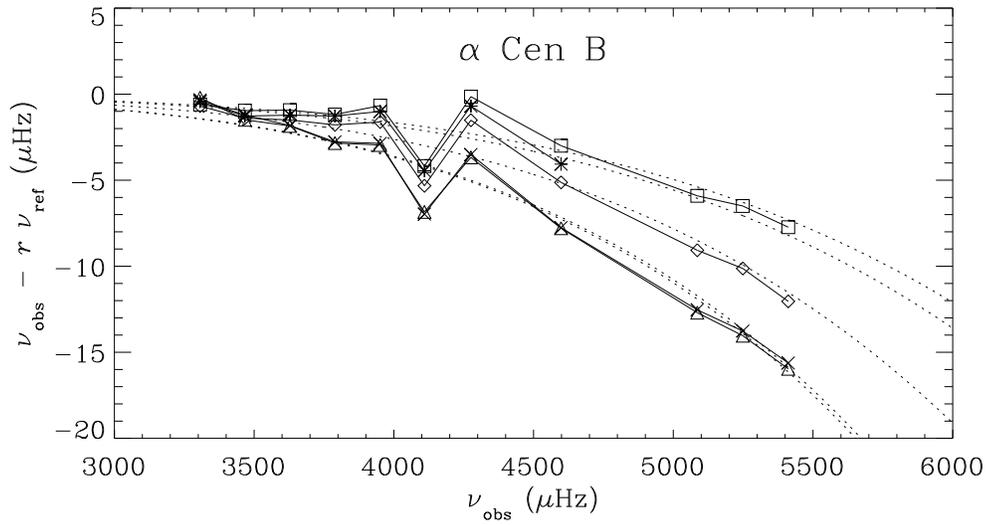}
\caption[]{\label{fig.acenb} Same as Fig.~\ref{fig.bhyi}, but for \acenb.
The models shown are: Model~B (squares), M2000~B (triangles), Model~S$^-$
(crosses), Th2002~B (asterisks) \new{and T2008 (diamonds)}.}
\end{figure}

For each model we used the value of $b$ found for the Sun, and used
equation~(\ref{eq.r}) to estimate $r$.  The results are shown in
Table~\ref{tab.calcs}, and also in Figs.~\ref{fig.bhyi}--\ref{fig.acenb}.
\new{Note that there is considerable scatter in the observed frequencies of
\acena{} and~B due to the relatively short span of the observations
relative to the mode lifetime.  Taking this into account, } it is once
again encouraging to see that the power law \new{(with a single value of
$b$)} provides a good fit to the frequency differences, and that for each
star there is good agreement between the densities derived from the
different models.  

\new{To estimate uncertainties, we have repeated the fits for the same
range of $b$ values considered in~\S\ref{sec.sun} (4.4--5.25) and again
found that there is no effect on the calculated density, to the precision
quoted in Table~\ref{tab.calcs}.  Over this range, the change in~$a$ is
less than 1\,\muHz{} for \bhyi{}, less than 0.4\,\muHz{} for \acena{} and
less than 0.7\,\muHz{} for \acenb.  The changes in the correction terms
plotted in Figs.~\ref{fig.bhyi}--\ref{fig.acenb} are comparable or slightly
bigger.  }

For each star, we can identify the model that is closest to being the
``best'' model as the one having $r$ closest to unity (note this is not
necessarily the model with the smallest near-surface offset).  These give
our best estimate of the stellar density for each star.
\myomit{and the corresponding frequency differences are plotted in
Fig.~\ref{fig.best}, this time as a function of normalized frequency
($\nu_{\rm obs}/\nu_0$).}

\citet{NDB2007} recently used interferometry to measure the angular
diameter of \bhyi{} to be $2.257\pm0.019$\,mas.  They combined this with
the parallax (from {\em Hipparcos}) and the mean density (from
asteroseismology) to determine the radius and mass of the star.  Using our
best estimate of the mean density of \bhyi{} ($0.258 \pm
0.001$\,g\,cm$^{-3}$) and the revised {\em Hipparcos} parallax ($134.070
\pm 0.110$\,mas; \citealt{vanLee2007}), we derive slightly updated values,
finding a radius of $1.809 \pm 0.015\,R_\sun$ (0.85\%) and a mass of $1.085
\pm 0.028\,M_\sun$ (2.6\%).

\section{Discussion and Conclusions}

The method outlined here for correcting near-surface effects can be applied
to model frequencies before they are compared with observations.  As in the
case of the Sun, we expect that the correction is independent of degree at
a given frequency, for low-degree acoustic modes, and thus the correction
determined from radial modes can be applied to all such modes.

There is, however, an important exception.  In evolved stars, mixed
modes may be observed that have the character of gravity modes in the deep
interior of the star.  Observational evidence for mixed modes has been
found in \eboo{} \citep{KBV95,KBB2003,CEB2005} and \bhyi{} \citep{BKA2007}.
Owing to the larger amplitude of these modes in the stellar interior and
hence their higher inertia, their frequencies are less affected by the
surface effects.  Techniques need to be developed to take this into account
in the application of the surface correction before the frequencies are
analyzed.

\new{The use of a single power law is made plausible by the knowledge that
the offsets we are modelling are caused by the properties of the
near-surface layers and hence presumably depend only on surface gravity,
effective temperature and composition, and not on the details of the
internal properties of the star.  This may be particularly true of the
exponent which, in the simple analysis by \citet{ChD+G80}, is determined by
an effective average polytropic index (i.e., the relation between pressure
and density) in the near-surface layers. }

It is important to note that, owing to its strong frequency dependence, the
offset also affects the large frequency separation $\Delta \nu$, as is
indeed implicit in equation~(\ref{eq.Deltanu}). \new{For example, although
Model~S is one of the best available models of the Sun, it has a large
separation that is 1\,\muHz{} greater than the observed value.}  Thus,
attempting to fit to stellar models based on $\Delta \nu$ will introduce
systematic errors, unless the corresponding correction is applied.

We have shown how to identify the model for each star that is closest to
the ``best'' model, by requiring that $r$ be as close as possible to unity.
This gives us an extremely accurate estimate of the mean stellar density.
However, it is important to point out that, while a model with $r$ close to
unity gives a good match to the overall structure of the star, it does not
necessarily reproduce the structure of the core or give a reliable estimate
of the stellar age.  Determination of those properties requires taking into
account the frequencies of the non-radial modes (including the small
frequency separations).  That is the next stage of model fitting, which can
be done after the surface correction has been applied to all modes.

\acknowledgments

We thank Teresa Teixeira for assistance in determining Models~A and~B, and
Mario Monteiro for providing the frequencies of Model~$S_0$ for \bhyi{} in
electronic form.  This work was supported financially by the Danish Natural
Science Research Council and the Australian Research Council.

\clearpage
\input{table1}
\input{table2}

\clearpage

\end{document}

%% file: table1.tex
\begin{deluxetable}{lcrrlcl}
\tablecolumns{6}
\tablewidth{0pc}
\tablecaption{Details of theoretical models \label{tab.models}}
\tablehead{
\colhead{Model ID} & \colhead{Star}   & \colhead{Mass}    & \colhead{Radius} &
\colhead{Luminosity} & \colhead{Age (Gyr)} & \colhead{Reference} }
\startdata
Model S     & Sun     & 1.000 & 1.000 & 1.000 &	4.52 & \citet{ChDDA96} \\
Model S$^-$ & \nodata & 1.000 & 0.930 & 1.286 &	2.25 & this paper \\
Model S$^+$ & \nodata & 1.000 & 1.128 & 0.832 &	7.44 & this paper \\
FM2003      & \bhyi   & 1.100 & 1.899 & 3.540 &	6.82 & Model $S_0$ from \citet{F+M2003} \\
Model H     & \bhyi   & 1.080 & 1.818 & 3.539 &	6.09 & this paper \\
Model H$^-$ & \nodata & 1.080 & 1.667 & 3.248 &	5.79 & this paper \\ 
Model H$^+$ & \nodata & 1.080 & 2.007 & 3.617 &	6.33 & this paper \\ 
Model~A     & \acena  & 1.111 & 1.224 & 1.5460&	6.96 & this paper \\
M2000~A     & \acena  & 1.160 & 1.228 & 1.527 &	2.71 & Model A$_{\rm BV}$ from \citet{MPL2000} \\
Th2002~A    & \acena  & 1.100 & 1.230 & 1.519 &	4.85 & \citet{TPM2002} \\
Model~B     & \acenb  & 0.928 & 0.867 & 0.5025&	6.88 & this paper \\
M2000~B     & \acenb  & 0.970 & 0.909 & 0.571 &	2.71 & Model B$_{\rm BV}$ from \citet{MPL2000} \\
Th2002~B    & \acenb  & 0.907 & 0.857 & 0.5002&	4.85 & \citet{TPM2002} \\
\new{T2008} & \acenb  & 0.929 & 0.869 & 0.4991&	5.86 & Model~M2 from \citet{TBG2008} \\
\enddata
\end{deluxetable}

%% file: table2.tex
\begin{table*}
\small
\begin{center}
\caption{\label{tab.calcs} Near-surface corrections}
\begin{tabular}{lccccccccc}
\tableline
\noalign{\smallskip}
\tableline
\noalign{\smallskip}
 & 
 & 
$\langle \nu_{\rm obs}(n)\rangle$ &
$\langle \Delta\nu_{\rm obs}(n)\rangle$ &
$\langle \nu_{\rm ref}(n)\rangle$ &
$\langle \Delta\nu_{\rm ref}(n)\rangle$ &
$a$ & 
 & 
$\bar{\rho}_{\rm ref}$
 & 
$r^2 \bar{\rho}_{\rm ref}$ \\
Model & 
$n_i$ & 
($\mu$Hz) &
($\mu$Hz) &
($\mu$Hz) &
($\mu$Hz) &
($\mu$Hz) &
$r$ & 
(g\,cm$^{-3}$) &
(g\,cm$^{-3}$) \\
\noalign{
\smallskip}
\tableline
\noalign{\smallskip}
\multicolumn{10}{c}{{Sun} ($\nu_0 = 3100$\,\muHz)}\\
\noalign{\smallskip}
Model S     & 17--25 & 3034.15 & 134.810 & 3038.95 & 135.854 &$ -4.73$ & 1.00000 &  1.408 &  1.408\\
Model~S$^-$ & 17--25 & 3034.15 & 134.810 & 3386.40 & 150.761 &$ -1.54$ & 0.89644 &  1.748 &  1.405\\
Model~S$^+$ & 17--25 & 3034.15 & 134.810 & 2540.67 & 114.155 &$ -8.67$ & 1.19770 &  0.982 &  1.408\\
\hline                                                                                              
\noalign{\smallskip}
\multicolumn{10}{c}{{\bhyi} ($\nu_0 = 1000$\,\muHz)}\\
\noalign{\smallskip}
Model H$^-$ & 13--19 & 1004.42 &  57.244 & 1134.75 &  65.985 &$ -4.57$ & 0.88978 &  0.329 &  0.260\\
Model H     & 13--19 & 1004.42 &  57.244 & 1001.90 &  58.417 &$ -5.19$ & 1.00847 &  0.253 &  0.258\\
FM2003      & 13--19 & 1004.42 &  57.244 &  948.58 &  55.714 &$ -6.90$ & 1.06723 &  0.226 &  0.258\\
Model H$^+$ & 13--19 & 1004.42 &  57.244 &  868.28 &  50.654 &$ -5.32$ & 1.16385 &  0.188 &  0.255\\
\hline
\noalign{\smallskip}
\multicolumn{10}{c}{{\acena} ($\nu_0 = 2400$\,\muHz)}\\
\noalign{\smallskip}
Model~S$^+$ & 16--26 & 2335.89 & 105.541 & 2509.96 & 114.026 &$ -3.18$ & 0.93196 &  0.982 &  0.853\\
M2000~A     & 16--26 & 2335.89 & 105.541 & 2362.88 & 107.949 &$ -6.48$ & 0.99141 &  0.882 &  0.867\\
Model~A     & 16--26 & 2335.89 & 105.541 & 2345.37 & 106.344 &$ -2.05$ & 0.99686 &  0.854 &  0.849\\
Th2002~A    & 16--26 & 2335.89 & 105.541 & 2303.36 & 104.927 &$ -4.78$ & 1.01626 &  0.833 &  0.860\\
\hline
\noalign{\smallskip}
\multicolumn{10}{c}{{\acenb} ($\nu_0 = 4100$\,\muHz)}\\
\noalign{\smallskip}
Th2002~B    & 19--27 & 3890.70 & 161.482 & 3910.71 & 162.605 &$ -2.11$ & 0.99534 &  2.030 &  2.011\\
Model~B     & 19--32 & 4261.05 & 161.988 & 4274.09 & 162.913 &$ -1.87$ & 0.99762 &  2.009 &  1.999\\
\new{T2008} & 19--32 & 4261.05 & 161.988 & 4261.92 & 162.697 &$ -2.96$ & 1.00087 &  1.995 &  1.998\\
M2000~B     & 19--32 & 4261.05 & 161.988 & 4064.00 & 155.400 &$ -4.15$ & 1.05006 &  1.819 &  2.006\\
Model~S$^-$ & 19--32 & 4261.05 & 161.988 & 3979.80 & 152.166 &$ -4.08$ & 1.07225 &  1.748 &  2.010\\
\noalign{
\smallskip}                             
\tableline                             
\end{tabular}
\end{center}
\end{table*}